# Phase convergence and crest enhancement of modulated wave trains

Hidetaka Houtani [1,*], Hiroshi Sawada [2] and Takuji Waseda [3]

[1]School of Engineering, The University of Tokyo, Bunkyo, Tokyo 113-8656, Japan
[2]Fluids Engineering and Ship Performance Evaluation Department, National Maritime Research Institute, Mitaka, Tokyo 181-0004, Japan
[3]Graduate School of Frontier Sciences, University of Tokyo, Kashiwa, Chiba 277-8563, Japan

*Correspondence: houtani@g.ecc.u-tokyo.ac.jp; 7-3-1 Hongo, Bunkyo, Tokyo, Japan

**Abstract:** The Akhmediev breather (AB) solution of the nonlinear Schrödinger equation (NLSE) shows that the maximum crest height of modulated wave trains reaches triple the initial amplitude as a consequence of nonlinear long-term evolution. Several fully nonlinear numerical studies have indicated that the amplification can exceed 3, but its physical mechanism has not been clarified. This study shows that spectral broadening, bound-wave production, and phase convergence are essential to crest enhancement beyond the AB solution. The free-wave spectrum of modulated wave trains broadens owing to nonlinear quasi-resonant interaction. This enhances bound-wave production at high wavenumbers. The phases of all the wave components nearly coincide at peak modulation and enhance amplification. We find that the phase convergence observed in linear-focusing waves can also occur due to nonlinear wave evolution. These findings are obtained by numerically investigating the modulated wave trains using the higher-order spectral method (HOSM) up to the fifth order, which allows investigations of nonlinearity and spectral bandwidth beyond the NLSE framework. Moreover, we confirm the crest enhancement through a tank experiment wherein waves are generated in the transition region from non-breaking to breaking. Owing to strong nonlinearity, the maximum crest height observed in the tank begins to exceed the HOSM prediction at an initial wave steepness of 0.10.

## 1. Introduction

Rogue waves (or freak waves) in the ocean can cause major damage to ships and offshore structures. The occurrence probability of such waves may need to be taken into account when designing or establishing rules for ships and offshore structures [1,2]. In addition to the occurrence probability, the shape of rogue waves affects the maximum wave load acting on ships [3]. The crest height of rogue waves is also of great concern for fixed offshore platforms because an air gap is the height between the wave crest and the platform deck [4,5].

Recent studies have revealed that modulational instability due to quasi-resonant interaction is one of the causes of rogue-wave formation [6-8]. Many studies have been conducted on the modulational instability or quasi-resonant interaction of water waves, starting from the discovery that Stokes waves are unstable under sideband modulations [9,10]. Such nonlinear wave evolutions are governed by the balance between nonlinearity and dispersion. This balance is expressed as a ratio between the wave steepness and spectral bandwidth (e.g., $\hat{\delta}$ for modulated wave trains [11] and the Benjamin–Feir index (BFI) for irregular waves [6]). The ratio $\hat{\delta}$ affects the initial growth of unstable sidebands [9], the recurrence period [12], and the maximum amplitude or crest height [13-16] of modulated wave trains. Moreover, the BFI is a key parameter for the occurrence probability of rogue waves [6,17].

This study focuses on the maximum crest height of modulated wave trains, or the maximum amplification of modulated wave trains, which is defined as the ratio between the maximum crest height and the initial Stokes-wave amplitude. This has been addressed in several experimental [18,19] and numerical [19-21] studies. Su and Green [18] and Tanaka [20] investigated the variation in maximum amplification against initial wave steepness. Through numerical simulation, Tanaka [20] showed that the maximum amplification predicted by the nonlinear Schrödinger equation (NLSE) [22] and a two-dimensional fully-nonlinear (FNL) potential flow solver [23] was much higher than the experimental results of Su and Green [18]. Tanaka's FNL simulation also showed that the maximum amplification could exceed 3 depending on the initial wave steepness. On the basis of experimental results and numerical results from the Dysthe equation [24], Waseda [19] showed that the difference between the numerical results of Tanaka [20] and the experimental results of Su and Green [18] could be explained, to a certain degree, by the influence of the spectral bandwidth of modulated wave trains. Su and Green [18] and Tanaka [20] determined that the spectral bandwidth had a one-to-one correspondence with the initial wave steepness. However, the maximum amplification can differ significantly depending on the spectral bandwidth for a given initial wave steepness. Slunyaev and Shrira [21] investigated the dependence of the maximum amplification of modulated wave trains on both the initial wave steepness and spectral bandwidth by analyzing the Akhmediev breather (AB) solution of the NLSE [13,25] and performing FNL simulation based on conformal mapping [26]. In this FNL simulation, a maximum amplification larger than 4 was observed in the specific case of a very narrow spectral bandwidth. Waseda [19] pointed out that, for a given initial wave steepness, the maximum amplification increases as the spectral bandwidth narrows. This relation was analytically explained by the AB solution in the cubic NLSE regime [14-16,21].

Such work has clarified the significant influence of the initial wave steepness and spectral bandwidth on the maximum crest height of modulated wave trains. However, it is still unclear why the maximum amplification of modulated wave trains can exceed 3, although AB predicts the maximum to be 3 in the limit of zero spectral bandwidth [14-16, 21]. Therefore, the aim of this study is to clarify the physics behind crest enhancement of modulated wave trains from the spectral-broadening and phase-convergence [27] perspectives. To investigate crest enhancement, we carried out tank experiments generating modulated wave trains and performed corresponding numerical simulations using the higher-order spectral method (HOSM) [28,29]. The evolution of the spectral broadening and degree of phase convergence were analyzed using the HOSM outputs in the non-wave-breaking regime. Note that the maximum wave height and maximum trough depth are parameters similar to the maximum crest height that this study focuses on. However, for a given modulated wave train, these three values differ [21] because of different bound-wave contributions to them [30].

In Section 2, we describe the set-up for the numerical simulations and experiments on modulated wave trains. The simulated and experimental results are compared in Section 3. We discuss the mechanism of crest enhancement of modulated wave trains in Section 4, and the conclusions of this study follow in Section 5.

## 2. Facility and Methods
*2.1. Numerical Simulations*

We numerically simulated the temporal evolution of spatially periodic deep-water modulated wave trains using the HOSM [28,29] in the same manner as in our preceding studies [31,32]. The HOSM numerically solves Laplace's equation ($\nabla^2 \phi = 0$) subject to nonlinear kinematic and dynamic free-surface boundary conditions with respect to the surface elevation $\zeta$ and velocity potential on the free surface, $\Phi^S(=\phi|_{z=\zeta})$ [22]:

$$\begin{cases} \zeta_t + \nabla_x \zeta \cdot \nabla_x \Phi^S - (1 + \nabla_x \zeta \cdot \nabla_x \zeta)W = 0, \\ \Phi_t^S + g\zeta + (1/2)\nabla_x \Phi^S - (1/2)(1 + \nabla_x \zeta \cdot \nabla_x \zeta)W^2 = 0, \end{cases} \quad (1)$$

where $\nabla_x = (\partial/\partial_x, \partial/\partial_y)$, $W = (\partial \phi/\partial z)|_{z=\zeta}$, and $g$ denotes the gravitational acceleration. $\phi$ is expanded in a perturbation series up to an arbitrary nonlinear order $M$. This study adopted $M = 5$. The spatial derivatives ($\nabla_x[\cdot]$) were solved in wavenumber space using the fast Fourier transform, which enabled efficient calculation. To remove spurious high-frequency waves arising from aliasing, the following low-pass filter proposed in [28] was applied:

$$|k| < \frac{N_x}{M+1} dk, \quad (2)$$

where $k$, $dk$, and $N_x$ denote the wavenumber, wavenumber interval, and number of spatial nodes in the $x$ direction, respectively. This study only addressed unidirectional modulated wave trains propagating in the $x$ direction. The HOSM cannot take into account wave breaking directly because the free surface $\zeta$ is assumed to be a single-valued function with respect to the horizontal coordinate $x$. However, when a wave sufficiently steep to break appears in the HOSM simulation, the low-pass filter in Eq. (2) removes the energies of high-wavenumber components. Accordingly, the computation can continue to some extent beyond possibly breaking events [33,34].

The initial wave profile of the HOSM simulation was a three-wave system consisting of a carrier, upper sideband, and lower sideband (denoted as $c$, $+$, and $-$, respectively):

$$\zeta(x) = a_c \cos(k_c x) + a_+ \cos(k_+ x + \psi_+) + a_- \cos(k_- x + \psi_-), \quad (3)$$

where $a$, $k$, and $\psi$ denote the amplitude, wavenumber, and phase, respectively. $k_\pm$ is defined as $k_c \pm \delta k$, where $\delta k$ is the perturbation wavenumber. Table 1 lists the parameters of the initial wave profiles of the modulated wave trains used in the HOSM simulation. We swept the initial wave steepness $a_0 k_c$ between 0.08 and 0.115 while the spectral bandwidth, that is, the perturbation wavenumber $\delta k/k_c$, was fixed. The critical parameter $\hat{\delta}$ introduced in Section 1, which governs the nonlinear evolution of modulated wave trains [11], is given by

$$\hat{\delta} = \frac{1}{2} \frac{\delta k/k_c}{a_0 k_c} \quad (4)$$

and was systematically varied from 0.62 to 0.89.

**Table 1.** Parameters defining the initial profiles of the modulated wave trains simulated with the HOSM. $a_0 \left[= (a_c^2 + a_+^2 + a_-^2)^{1/2}\right]$ denotes the initial amplitude of the Stokes wave.

| Parameters | Values |
| --- | --- |
| wavelength $\lambda_c (= 2\pi/k_c)$ | 3 m |
| perturbation wavenumber $\delta k/k_c$ | 1/7 |
| sideband wave amplitudes $a_\pm/a_0$ | 0.1 |
| sideband wave phases $\varphi_\pm$ | $-\pi/4$ |
| wave steepness $a_0 k_c$ | 0.08 – 0.115 |
| $\hat{\delta}$ | 0.62 – 0.89 |

The initial wave profile [Eq. (3)] was given on the basis of linear wave theory and did not satisfy the fully nonlinear free-surface boundary condition [Eq. (1)]. Therefore, we adopted the nonlinear wave-initialization method proposed in [35] (see also [32]). The initially linear wave field was gradually transformed into a nonlinear wave field with an adjustment period $T_a$. $T_a = 32 T_c$, where $T_c$ denotes the period of the carrier wave.

*2.2. Tank Experiment*

We performed a wave-generation experiment in a wave tank (WT) (50 m × 8 m × 4.5 m) (Figure 1) at the National Maritime Research Institute to compare its results with those of the HOSM simulation and to investigate modulated wave trains including wave breaking. We generated the modulated wave trains using the HOSM wave generation (HOSM-WG) method [32]. A nonlinear wave field precomputed with the HOSM was generated in a wave tank by sending a temporally evolving signal calculated from the HOSM output to a wave maker. HOSM-WG can control when and where the maximum crest height appears in a wave tank. In this study, we generated the modulated wave trains such that the maximum crest appeared at $t = 40$ s after the beginning of wave generation and at $x = 12$ m from the wave maker in the WT.

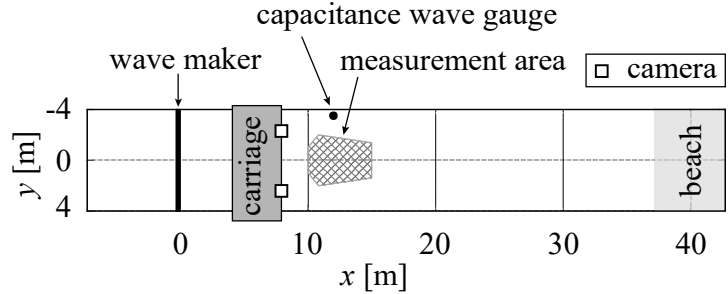

**Figure 1.** Schematic of the stereo camera set up in the wave tank (WT).

A capacitance wave gauge was set at $x = 12$ m to measure the wave-elevation time series. However, the location of the maximum crest could deviate from $x = 12$ m, especially when wave breaking occurred. To measure the maximum crest height even in cases of wave breaking, we measured the wave-surface profiles using a stereo-imaging technique [36,2]. About 100 spherical floats with a diameter of less than 20 mm were set on the wave surface, and two cameras tracked the three-dimensional motion of these floats. We acquired the wave profiles by fitting a smoothing spline curve to the floats' three-dimensional coordinates [2] and evaluated the maximum crest height. The set-up of the stereo cameras in the WT is illustrated in Fig. 1, where the measurement area is also indicated. The estimated error in the crest height of regular waves with a wavelength of 3 m and wave height between 10 and 20 cm is less than 4% in this stereo-imaging scheme [36]. Note that the standard deviation of the wave-maker motion was found to be 1.065 times larger than the given signal owing to a problem with controlling the mechanical wave maker [31]. Therefore, the experimental results presented in Section 3 are corrected by a factor of 1.065 for comparison with the HOSM simulation.

## 3. Results of Numerical Simulations and Experiments

In this section, we compare the maximum crest height of the modulated wave trains in the HOSM simulation and the HOSM-WG experiment in the WT. Figure 2 presents the variation in the normalized maximum crest height $\zeta_{cr}/a_0$ with the initial wave steepness $a_0 k_c$. For reference, Figure 2 also shows the predictions of the AB solution of the NLSE with and without the second-order bound-wave correction. The maximum crest height of the free wave, $\zeta_{cr}^{(AB;f)}$, and that taking into account the bound wave, $\zeta_{cr}^{(AB;f+b)}$, for the AB solution are given respectively by

$$\frac{\zeta_{cr}^{(AB;f)}}{a_0} = 1 + 2\sqrt{1 - \frac{1}{2}\hat{\delta}^2} \qquad (5)$$

and

$$\frac{\zeta_{cr}^{(AB;f+b)}}{a_0} = \frac{\zeta_{cr}^{(AB;f)}}{a_0}\left\{1 + \frac{1}{2}k_c\zeta_{cr}^{(AB;f)}\right\}. \qquad (6)$$

These are derived in Appendix A. The definition of $\hat{\delta}$, expressing the balance between the initial wave steepness and spectral bandwidth, is given in Eq. (4). The quantity $\zeta_{cr}^{(AB;f)}/a_0$ reaches a maximum of 3 in the limit $\hat{\delta} \to 0$.

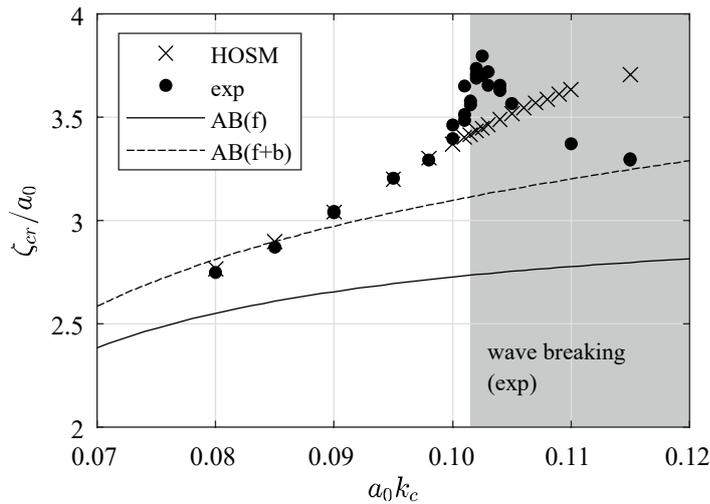

**Figure 2.** Variation in maximum crest height with the initial wave steepness. $f$ and $b$ indicate the free and second-order bound waves, respectively.

Overall, $\zeta_{cr}/a_0$ increases with $a_0 k_c$ both in the WT experiment and HOSM simulation. These values are notably larger than the free-wave AB prediction [Eq. (5)]. This gap is compensated by adopting the second-order AB prediction [Eq. (6)] at low $a_0 k_c$, which indicates a substantial contribution of the bound waves to the maximum crest height. However, the results of the HOSM simulation and WT experiments are still larger than this second-order AB prediction at high $a_0 k_c$ ($a_0 k_c > 0.090$). Moreover, the experimental results begin to deviate from the HOSM results at approximately $a_0 k_c = 0.100$. The experimental value of $\zeta_{cr}/a_0$ starts to decrease at $a_0 k_c = 0.1025$, while $\zeta_{cr}/a_0$ continues to increase in the HOSM simulation. This deviation can be attributed to stronger nonlinearity in the WT experiment. Wave breakings were observed visually in the WT experiment for $a_0 k_c > 0.1015$ (indicated with a gray shade in Figure 2), although wave breaking could not be reproduced in the HOSM simulation as explained in Section 2.1. This stronger nonlinearity led to a higher crest height at approximately $a_0 k_c = 0.1015$ in the experiment. Beyond the breaking/non-breaking margin ($a_0 k_c = 0.1015$), the maximum crest height $\zeta_{cr}/a_0$ decreased with $a_0 k_c$ because larger wave breakings occurred prior to the modulation peak.

We compare these frequency spectra in Figure 3 to clarify the cause of the differences in crest height of the modulated wave trains among the WT experiment, HOSM simulation, and AB predictions. The spectra in Figure 3 were evaluated from the wave-elevation time series covering one wave-group period at approximately the time of the maximum crest height ($t = t_{max}$). Figure 3(a) presents the frequency spectra of the modulated wave train with $a_0 k_c = 0.100$, in which the maximum crest heights in the WT experiment and HOSM simulation almost agree but are notably larger than the second-order AB prediction. The wave spectra in the WT experiment and HOSM simulation are broader than those in the AB predictions. The energy difference is significant especially at $\omega/\omega_c > 2.5$ because the AB solutions consider the bound-wave contribution up to the second order. Meanwhile, substantial differences in the spectra near the peak frequency ($\omega/\omega_c = 1$) are also observed. The energy in the WT and HOSM results at approximately $\omega/\omega_c = 1.5$ is higher than the AB prediction, while the energy in the WT and HOSM results at approximately $\omega/\omega_c = 0.7$ is lower than the AB prediction. The spectral difference around the peak frequency indicates the difference in free-wave spectral evolution due to quasi-resonant interaction. The NLSE, the governing equation of the AB solution, assumes a narrow-band spectrum, while the WT experiment and HOSM simulation do not restrict the spectral bandwidth. We will demonstrate that the free-wave spectral broadening is larger in the HOSM than in the AB solution in Section 4.1.

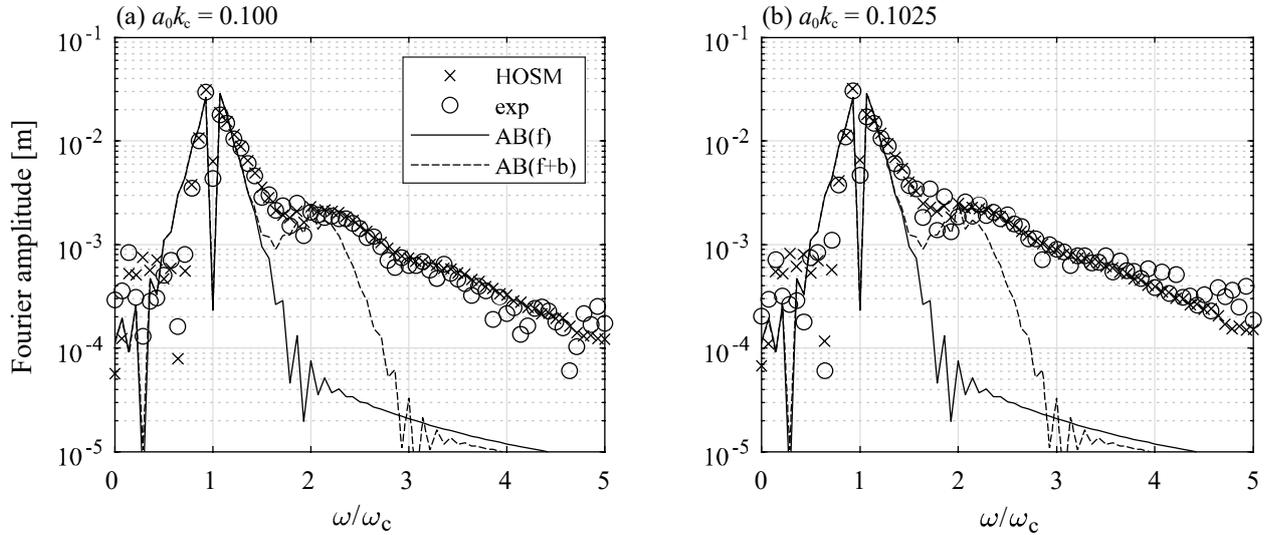

**Figure 3.** Comparison of the Fourier amplitudes of the modulated wave trains in the frequency domain among the experiment, HOSM simulation, and AB solution for (a) $a_0 k_c = 0.100$ (non-breaking) and (b) $a_0 k_c = 0.1025$ (slight breaking).

The wave spectrum with $a_0 k_c = 0.1025$ [Figure 3(b)], in which slight wave breaking was observed in the WT experiment, is almost the same as with $a_0 k_c = 0.100$ [Figure 3(a)]. A slight difference is observed at high frequencies ($\omega/\omega_c > 3.5$); the energy is slightly higher in the WT experiment than in the HOSM simulation. The larger maximum crest height in the WT experiment than in the HOSM simulation at $a_0 k_c = 0.1025$ can be interpreted as a consequence of the higher energy production at high frequencies in the WT experiment at $t = t_{max}$ due to strong nonlinearity. Whether the low energy at high frequencies can contribute to the maximum crest height will be discussed using the HOSM output in Section 4.1. We should note that the higher spectral energy in the WT experiment at $\omega/\omega_c > 3.5$ with $a_0 k_c = 0.1025$ may also be attributed to the high-frequency wave generation resulting from wave breaking. The wave-frequency spectrum in the WT result in Figure 3(b) includes the wave information not only at the instant $t = t_{max}$ but also over one wave-group period.

## 4. Discussion

### 4.1. Spectral Broadening and its Influence on Maximum Crest Height

The comparison of the WT experimental results with the HOSM simulation and AB solutions in Section 3 implies the spectral broadening and the bound waves influence the maximum crest height of modulated wave trains. Therefore, in this section, we discuss these influences by scrutinizing the HOSM output. The discussion in the following subsections is based on the HOSM simulation results and specifically confined to a non-breaking potential-flow regime.

We begin by investigating the spectral broadening of modulated wave trains during their nonlinear evolution using HOSM outputs. Figure 4 presents the temporal evolution of the wave profile and wavenumber amplitude spectrum of a modulated wave train with $a_0 k_c = 0.105$. The figure only depicts the period around the time of the maximum crest height ($t = t_{max}$), from $25T_c$ before to $15T_c$ after $t = t_{max}$. The wavenumber spectrum broadens before the crest height reaches its maximum ($t < t_{max}$), becomes broadest at approximately the time of the maximum crest height ($t = t_{max}$), and then narrows afterward.

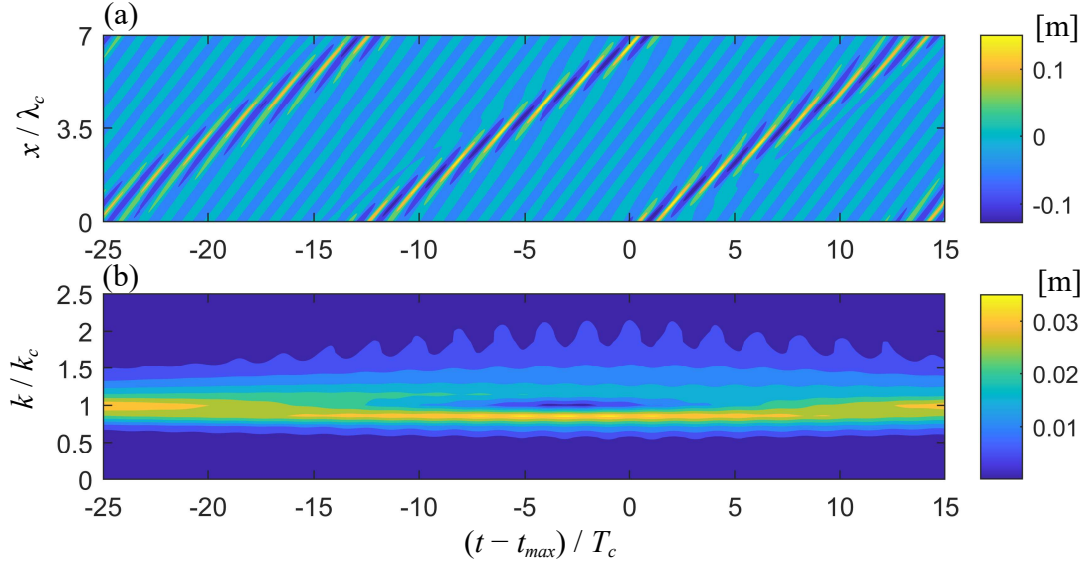

**Figure 4.** Temporal evolution of the modulated wave train with $a_0 k_c = 0.105$ near the time of the maximum crest height ($t = t_{max}$) in the HOSM simulation. (a) Wave elevation $\zeta(x,t)$; (b) amplitude spectrum $|\hat{\zeta}(k,t)|$.

To quantify the spectral broadening, we introduce the indicator $\Delta K$, defined as the mean wavenumber difference from the carrier wavenumber ($\Delta k = k - k_c$) weighted by the Fourier amplitude [37,38]:

$$\Delta K = \left| \frac{\sum_j \Delta k_j^2 |\hat{\zeta}(k_j)|^2}{\sum_j |\hat{\zeta}(k_j)|^2} \right|^{1/2}, \text{ with } \Delta k_j = k_j - k_c. \qquad (7)$$

Here, $j$ and $\hat{\zeta}(k)$ denote the wavenumber component and complex Fourier amplitude of a wave train in wavenumber space, respectively, and $\Sigma_j$ expresses the sum over all wavenumber components. Figure 5 shows the temporal evolution of the normalized mean wavenumber difference $\Delta K/\delta k$ of the modulated wave train with $a_0 k_c = 0.105$. The temporal evolution of the wavenumber spectrum [Figure 4(b)] indicates that $\Delta K$ reaches its maximum at $t = t_{max}$. If all the energy is transferred only to the sideband waves ($k = k_\pm$), $\Delta K/\delta k$ becomes 1. Therefore, the maximum value of $\Delta K/\delta k = 3.64$ at $t = t_{max}$ indicates that the energy is transferred further beyond the sideband wavenumber components at approximately $t = t_{max}$. Note that $\Delta K/\delta k$ is 0.141 in the initial state ($t = t_{ini}$), in which the initial wave profiles are given as a three-wave system [Eq. (3) and Table 1]. This value is also indicated by a dashed line in Figure 5.

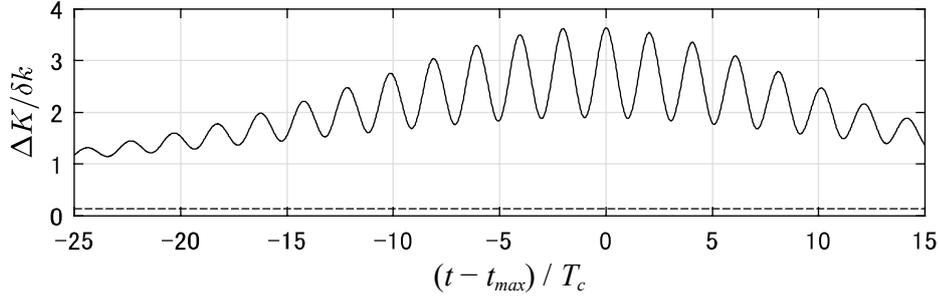

**Figure 5.** Temporal evolution of the mean wavenumber difference $\Delta K/\delta k$ of the modulated wave train with $a_0 k_c = 0.105$ at approximately the time of the maximum crest height ($t = t_{max}$) in the HOSM simulation. The dashed line represents the initial value.

To investigate the dependence of the spectral broadening at $t = t_{max}$ on the initial wave steepness, or $\hat{\delta}$ [Eq. (4)], we plot $\Delta K/\delta k$ at $t = t_{max}$ in the HOSM simulation against $a_0 k_c$ in Figure 6. The spectral broadening $\Delta K/\delta k$ at $t = t_{max}$ increases as the wave steepness $a_0 k_c$ increases. Moreover, the spectral broadening is larger than that predicted by AB (solid line), which takes into account the second-order bound waves (Appendix A). The deviation becomes larger as $a_0 k_c$ increases. From the definition of $\Delta K$ [Eq. (7)], the deviation of $\Delta K/\delta k$ is conjectured to be mainly due to the difference in energy at high wavenumbers far from $k = k_c$. Indeed, notable deviations in spectral energy between the HOSM and AB are observed at high frequencies in Figure 3. The bound-wave energy is dominant at such high wavenumbers (or high frequencies), as demonstrated next.

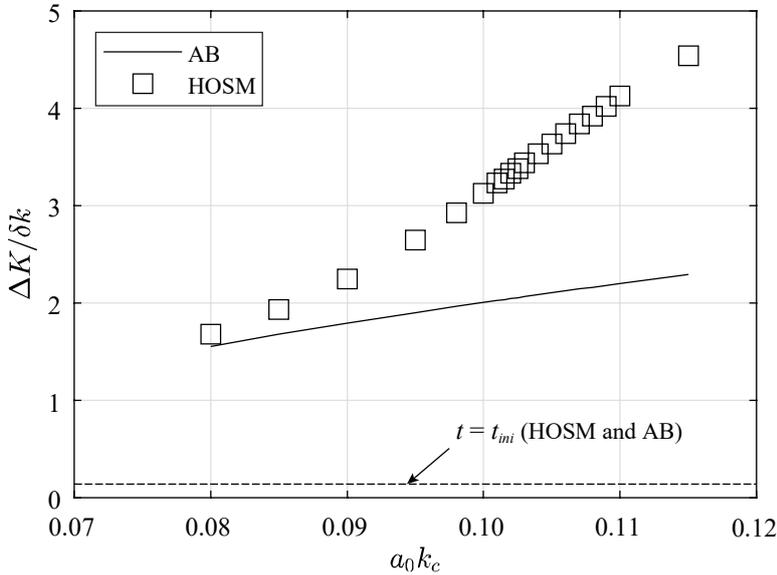

**Figure 6.** Relation between the initial wave steepness $a_0 k_c$ and mean wavenumber difference $\Delta K/\delta k$ at $t = t_{max}$ for the modulated wave train. The dashed line represents the initial value.

To clarify the origin of the difference in spectral broadening between the HOSM and AB observed in Figure 6, we next investigate the spectral broadening for free- and bound-wave components individually. For this purpose, we separated them by applying an ideal filter to the wavenumber–frequency spectrum of the HOSM outputs [31,32]. We obtained the total, free-wave, and bound-wave amplitude spectra at $t = t_{max}$ as shown in Figure 7 and evaluated $\Delta K$ individually from these spectra. We define $\Delta K$ for free and bound waves ($\Delta K^{(f)}$ and $\Delta K^{(b)}$) as follows:

$$\Delta K^{(f,b)} = \left| \frac{\sum_j \Delta k_j^2 |\hat{\zeta}^{(f,b)}(k_j)|^2}{\sum_j |\hat{\zeta}^{(t)}(k_j)|^2} \right|^{1/2}, \quad \text{with} \quad \Delta k_j = k_j - k_c. \tag{8}$$

The superscripts $t$, $f$, and $b$ denote the total, free, and bound waves, respectively.

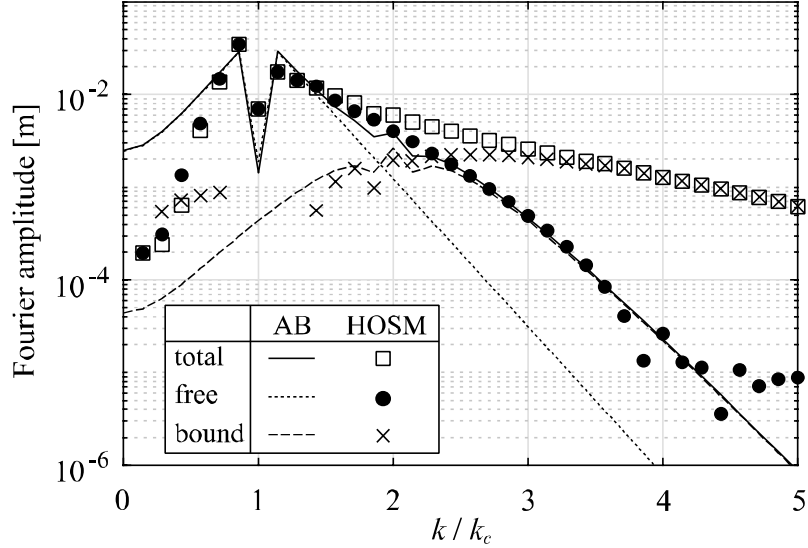

**Figure 7.** Comparison of the Fourier amplitudes of the modulated wave trains with $a_0 k_c = 0.105$ in the wavenumber domain at the time of the maximum crest height in the HOSM simulation and AB solution. Note that the bound-wave spectrum of the HOSM simulation breaks around $k/k_c$ because its spectral energy is removed owing to the ideal filter separating free and bound waves.

The relation between $a_0 k_c$ and $\Delta K / \delta k$ is presented in Figure 8(a) for the free waves and Figure 8(b) for the bound waves. The difference between the HOSM and AB is greater for the bound waves than for the free waves. This confirms that the primary cause of the difference in total spectral broadening between the HOSM and AB in Figure 6 is the difference in bound wave energy. As conjectured, a significant energy difference is observed between the HOSM and AB at high wavenumbers (Figure 7). The reason bound-wave production at high wavenumbers is more energetic in the HOSM than in AB is the larger free-wave spectral broadening in the HOSM, as will be discussed next. This is because the bound waves are produced deterministically from the free-wave spectrum. Of course, bound waves higher than the second harmonics, not considered in the AB solution, also contribute to the higher bound-wave energy at high wavenumbers in the HOSM results. We could evaluate the bound waves correctly up to the fifth order in the HOSM simulation because this study adopted the nonlinear order $M = 5$.

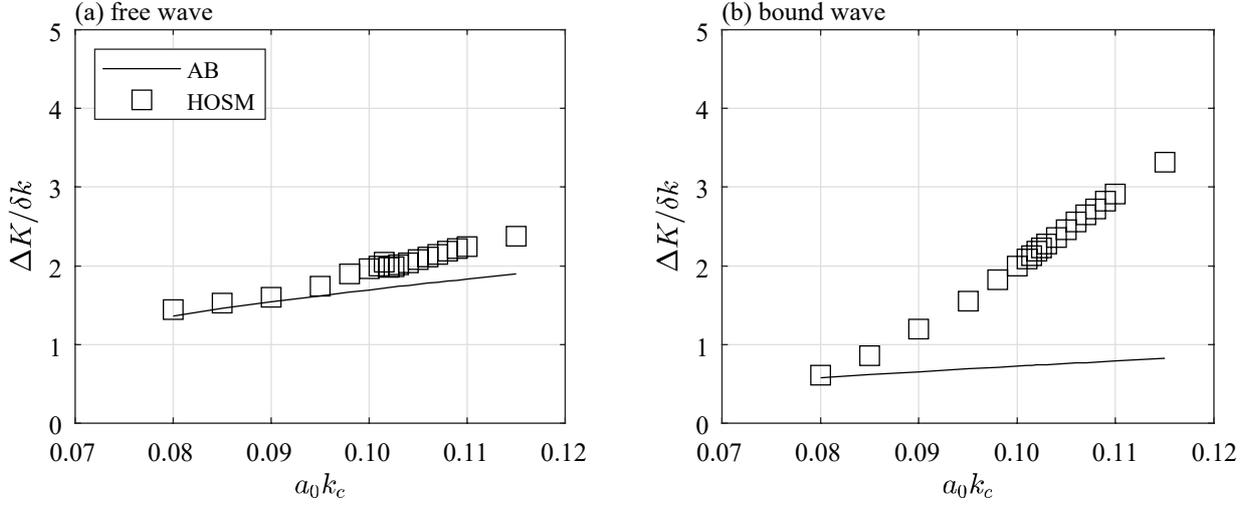

**Figure 8.** Relation between the initial wave steepness $a_0 k_c$ and mean wavenumber difference $\Delta K/\delta k$ at $t = t_{max}$ for the modulated wave train. (a) Free wave; (b) bound wave.

As implied by the relation between $a_0 k_c$ and $\Delta K/\delta k$ for bound waves, free-wave spectra broaden further as $a_0 k_c$ increases [Figure 8]. The deviation in spectral broadening between the HOSM and AB becomes larger as $a_0 k_c$ increases. This deviation reflects the difference in the free-wave spectral shape at $t = t_{max}$ as indicated in Figure 7. The free-wave spectrum can broaden more in the HOSM than in AB because of the different treatments of the spectral bandwidth. The HOSM does not restrict the bandwidth, while the NLSE assumes narrow-banded wave spectra. From the investigation so far, we can conclude that the free-wave spectral broadening and resultant bound-wave production of the modulated wave trains at high wavenumbers result in larger spectral broadening in the HOSM than in AB. Janssen [6] has observed a similar relation between the wave steepness and spectral broadening in irregular waves. He has shown that the spectra of irregular waves broaden as the ratio of the wave steepness to the spectral bandwidth increases, owing to the enhanced quasi-resonant interaction.

Next, we interpret the relation between the spectral broadening and crest enhancement by introducing the "amplitude sum" [39]

$$A_s = \sum_j |\hat{\zeta}(k_j)|. \tag{9}$$

$A_s$ is defined as the sum of the Fourier amplitudes of all the spectral components and, accordingly, expresses the potential maximum crest height when all the wave components are in phase. Furthermore, $A_s$ generally increases as the energy spreads over more wave components in a system in which the total wave energy $\left(E = \sum_j |\hat{\zeta}(k_j)|^2\right)$ is conserved [39].

Figure 9 presents the temporal evolution of the normalized amplitude sum $A_s/a_0$ of the modulated wave train with $a_0 k_c = 0.105$ in the HOSM simulation. $A_s/a_0$ temporally varies and reaches its maximum at approximately $t = t_{max}$, which is similar to the temporal evolution of $\Delta K/\delta k$ (Figure 5). From this temporal evolution of $A_s/a_0$ and that of $\Delta K/\delta k$, we can conclude that the potential maximum crest height increases according to the spectral broadening during the nonlinear wave evolution. It is interesting that the time of the maximum $A_s$ (indicated with a triangle in Figure 9) does not coincide precisely with $t = t_{max}$. $A_s$ reaches its maximum a few wave periods after $t = t_{max}$. This time lag will be discussed in Section 4.3.

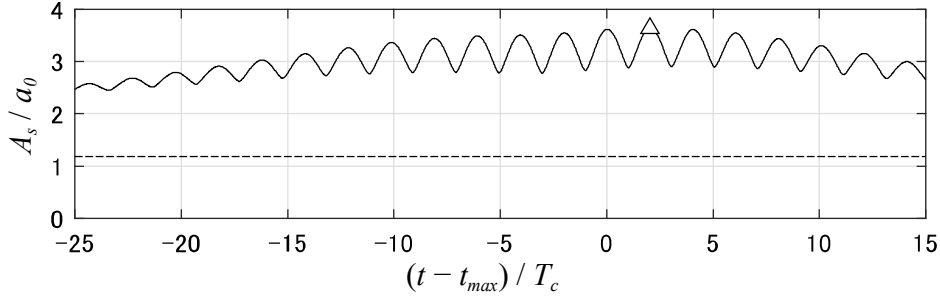

**Figure 9.** Temporal evolution of the amplitude sum $A_s$ of the modulated wave train with $a_0 k_c = 0.105$ at approximately the time of the maximum crest height ($t = t_{max}$) in the HOSM simulation. The dashed line represents the initial value, and the triangle indicates the maximum value.

Figure 10(a) presents the relation between the spectral broadening $\Delta K/\delta k$ and amplitude sum $A_s/a_0$ in the HOSM simulation. $A_s/a_0$ at $t = t_{max}$ increases as $\Delta K/\delta k$ increases. The potential maximum crest height of the modulated wave train increases as the spectrum broadens. For a given $\Delta K/\delta k$, $A_s/a_0$ is lower than the AB prediction (solid curve). However, the range of the spectral broadening $\Delta K/\delta k$ notably differs between the HOSM and AB (1.56 < $\Delta K/\delta k$ < 2.29 for AB and 1.68 < $\Delta K/\delta k$ < 4.54 for the HOSM) for the $a_0 k_c$ range investigated here ($0.08 \leq a_0 k_c \leq 0.115$). Therefore, with larger spectral broadening than in AB, the modulated wave train in the HOSM attains an $A_s/a_0$ value exceeding the maximum AB prediction ($A_s/a_0 = 3.50$) at $\Delta K/\delta k > 3.38$.

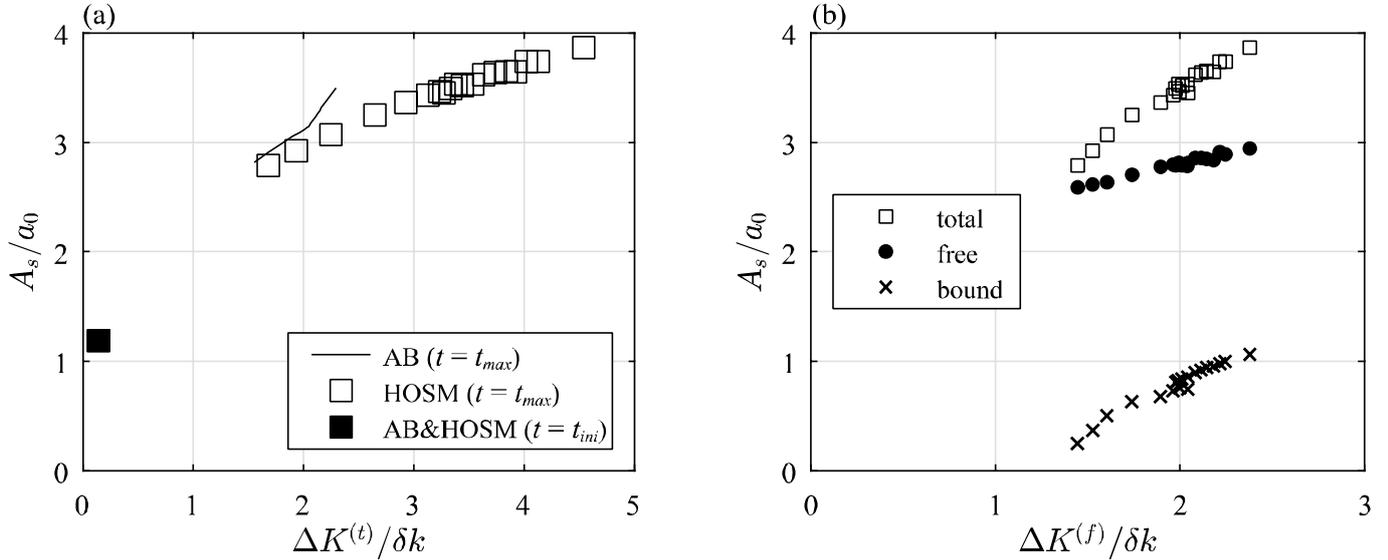

**Figure 10.** (a) Relation between the spectral broadening $\Delta K/\delta k$ and amplitude sum $A_s/a_0$ of the modulated wave train. (b) Relation between the free-wave spectral broadening $\Delta K^{(f)}/\delta k$ and amplitude sum $A_s/a_0$ of the modulated wave train for the total, free-wave, and bound-wave components.

To clarify the individual contributions of each wave type to $A_s$, we plot $A_s/a_0$ for the free and bound waves together with that for the total wave against $\Delta K/\delta k$ in Figure 10(b). Unlike Figure 10(a), Figure 10(b) uses $\Delta K/\delta k$ for the free wave ($\Delta K^{(f)}/\delta k$) because the free-wave spectral broadening governs the bound-wave and total-wave spectral broadening. $A_s/a_0$ for the free wave is observed to increase as the free-wave spectrum broadens. However, for the modulated wave trains we have investigated, $A_s/a_0$ for the free waves does not exceed 3, which is the maximum amplification predicted by the free-wave AB solution. This means that an amplification larger than 3 is never achieved only from free-wave spectral broadening, but is achieved with contributions from bound waves. The contribution of the bound waves to $A_s/a_0$ becomes larger as $a_0 k_c$ increases. The bound-wave contribution to $A_s$ increases from 8.7% for $\Delta K^{(f)}/\delta k = 1.45$ ($a_0 k_c = 0.08$) to 27% for $\Delta K^{(f)}/\delta k = 2.38$ ($a_0 k_c = 0.115$). This result indicates that energized bound-wave production at high wavenumbers is a consequence of free-wave spectral broadening, as discussed earlier in this section (Figures 7 and 8), and is crucial for crest enhancement of modulated wave trains. We should note that $A_s(\text{free}) + A_s(\text{bound})$ does not necessarily coincide with $A_s(\text{total})$ because $|\hat{\zeta}(k)^{(\text{total})}| \neq |\hat{\zeta}(k)^{(\text{free})}| + |\hat{\zeta}(k)^{(\text{bound})}|$ when the phases of the free and bound waves do not coincide. Of course, $\hat{\zeta}(k)^{(\text{total})} = \hat{\zeta}(k)^{(\text{free})} + \hat{\zeta}(k)^{(\text{bound})}$ holds at any time. Therefore, the bound-wave contribution to $A_s$ was evaluated as the ratio of $A_s(\text{bound})$ to $A_s(\text{free}) + A_s(\text{bound})$ here.

As stated in Section 1, bound waves have different contributions to the maximum crest height and the maximum trough depth. Contrary to the crest-height amplification examined above, energized bound-wave production contributes to trough-depth suppression for modulated wave trains in analogy with Stokes wave theory [30]. Therefore, free-wave spectral broadening should enhance crest height–trough depth asymmetry. FNL simulation of modulated wave trains by Slunyaev and Shrira [21] has indicated such crest height–trough depth asymmetry. They have observed that the difference between the maximum crest height and the maximum trough depth becomes more prominent as the wave steepness increases.

In Section 3, we pointed out the possibility that the reason the maximum crest height in the WT experiment is larger than in the HOSM simulation at approximately $a_0 k_c = 0.1025$ is that the spectral energy is higher at high wave frequencies ($\omega/\omega_c > 3.5$). To analyze this possibility, we evaluated the contribution of high-wavenumber components to $A_s$. We found that the contribution of components with $k/k_c > 3.5$ to the maximum crest height of the modulated wave trains with $a_0 k_c = 0.1025$ was 7.4%. The Fourier amplitudes of components with $k/k_c > 3.5$ are very small [$\sim O((a_0 k_c)^2)$] compared with the maximum Fourier amplitude at the lower sideband $k = k_-$ (Figures 3 and 7). However, the sum of the Fourier amplitudes for components with $k/k_c > 3.5$ is never infinitesimal and might contribute to the crest enhancement.

*4.2. Phase Convergence during Nonlinear Evolution of a Modulated Wave Train*

In Section 4.1, we found that the potential maximum crest height of the modulated wave train increased as the initial wave steepness increased. However, this result does not necessarily indicate an increase in the maximum crest height. Convergence of the phases of all the wave components [27] is necessary to achieve a crest height $\zeta_{cr}$ close to its potential maximum $A_s$. In a framework of linear wave superpositions, phase convergence is key to generating focusing waves [27,40]. Slunyaev and Shrira [21] showed that the phases of all the spectral components are nearly coincident in the AB solution. Therefore, in this section, we investigate the degree of phase convergence at the location and instant of the maximum crest height beyond the cubic NLSE regime using the HOSM output.

We define $x_{max}$ as the location of the maximum crest height at each time $t$. Thus, the maximum crest height at time $t$ is

$$\zeta_{cr} = \zeta(x_{max}, t) = \sum_j \text{Re}[\alpha(k_j)] \quad \text{with} \quad \alpha(k_j) \equiv \hat{\zeta}(k_j, t) \exp(ik_j x_{max}). \quad (10)$$

The modulus and argument of $\alpha_j$ express the Fourier amplitude and phase of the component waves at $x = x_{max}$ at time $t$, respectively. The phase $\varphi\ [\equiv \arg(\alpha)]$ is defined such that it becomes 0 when the crest of each wave component is at $x = x_{max}$.

Figure 11 presents an example of the amplitude and phase of the component waves at $t = t_{max}$ for $a_0 k_c = 0.105$. As observed in the phase spectrum [Figure 11(b)], most spectral components except the carrier wave at $k/k_c = 1$ are in phase at 0, especially free-wave components. This feature of phase convergence corresponds to the AB solution with a slight discrepancy. At the modulation peak, all of the AB free-wave components are in phase except for the carrier wave, which is in counter-phase with other components [21]. In this HOSM simulation, the carrier wave is out of phase with sideband waves but not counter-phase.

We also observe that the phases of some components with lower and higher wavenumbers ($k/k_c < 0.5$ and $k/k_c > 4$) are not necessarily 0. Second-order wave theory (Appendix B) explains that the subharmonic bound waves at low wavenumbers are in counter-phase with free waves. In addition, the out-of-phase components at high wavenumbers ($k/k_c > 4$) consist of free waves, and their energies are very low compared with those of bound waves at the same wavenumbers. Therefore, the contribution of such out-of-phase components at $k/k_c > 4$ to the crest height is considered almost negligible.

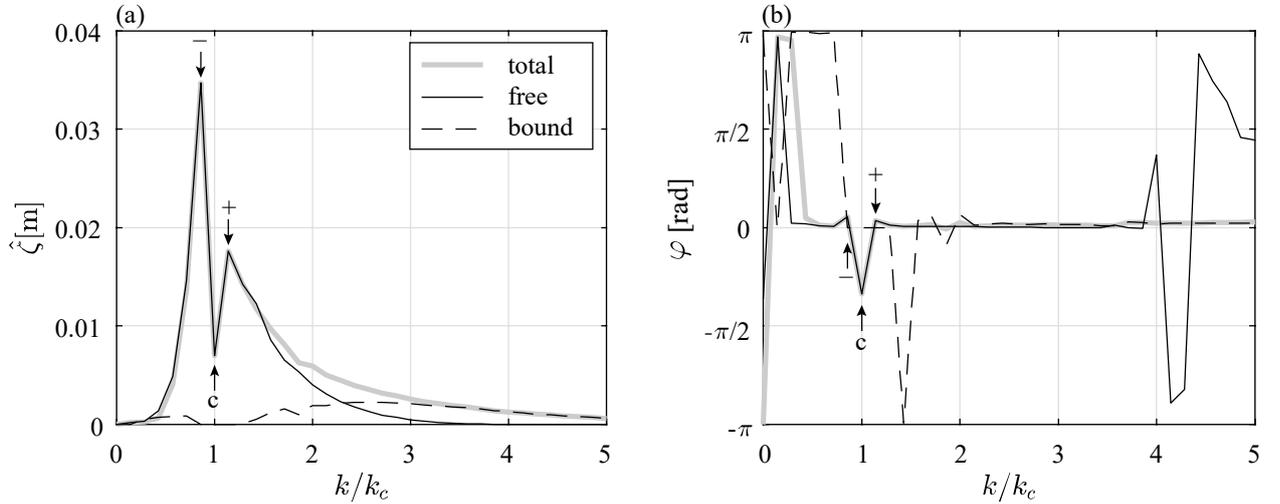

**Figure 11.** (a) Amplitude and (b) phase of the component waves of the modulated wave train with $a_0 k_c = 0.105$ at the location and time of the maximum crest height.

To quantify the degree of phase convergence, we introduce the parameter $D$ expressing the mean of $\cos \varphi(k_j)$ weighted by the Fourier amplitude $|\alpha(k_j)|$:

$$D = \frac{\sum_j |\alpha(k_j)| \cos \varphi(k_j)}{\sum_j |\alpha(k_j)|} \left( = \frac{\sum_j \mathrm{Re}[\alpha(k_j)]}{\sum_j |\alpha(k_j)|} = \frac{\zeta_{cr}}{A_s} \right). \qquad (11)$$

The definition of $\alpha$ is given in Eq. (10). As indicated in Eq. (11), the parameter $D$ can also be regarded as the ratio of the crest height $\zeta_{cr}$ to its potential maximum $A_s$. Figure 12 presents the temporal evolution of $D$ for the modulated wave train with $a_0 k_c = 0.105$. Similarly to the evolutions of $\Delta K$ and $A_s$, $D$ gradually increases when $t < t_{max}$, almost reaches 1 at approximately $t = t_{max}$, and starts to decrease afterward ($t > t_{max}$). This temporal evolution of $D$ confirms almost perfect phase convergence ($D \approx 1$) near $t = t_{max}$. It is interesting to note that the times of the maximum $D$ (indicated with a triangle in Figure 12) and $t = t_{max}$ do not coincide precisely. Unlike $A_s$, which reaches its maximum a few wave periods after $t = t_{ma}$ (Section 4.1), $D$ reaches its maximum a few wave periods ahead of $t = t_{max}$. This time difference will also be discussed in Section 4.3.

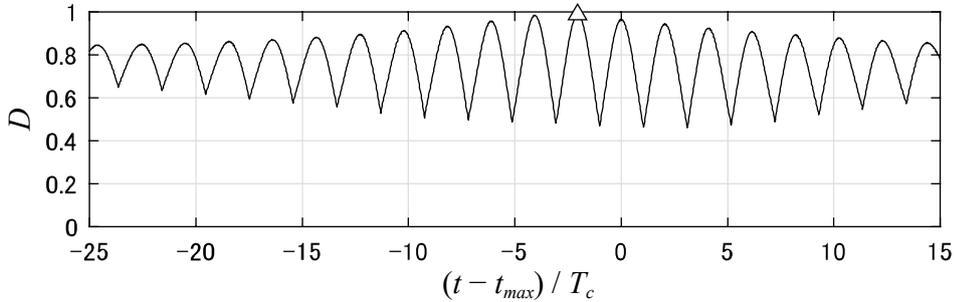

**Figure 12.** Temporal evolution of the phase convergence $D$ of the modulated wave train with $a_0 k_c = 0.105$ at approximately the time of the maximum crest height ($t = t_{max}$) in the HOSM simulation. The triangle indicates the maximum value.

To investigate the degree of phase convergence at $t = t_{max}$, we present $D$ at $t = t_{max}$ for all cases with $a_0 k_c$ from 0.08 to 0.115 in Figure 13. $D$ is close to 1 for all cases, although $D$ decreases slightly with increasing $a_0 k_c$. In other words, the phases of all the components nearly coincide at $t = t_{max}$, and result in a crest height exceeding 95% of its potential maximum $A_s$ for all cases. This result confirms that the near coincidence of the phases of all the spectral components observed in linear focusing waves [27,40] and the free-wave AB solution [21] holds well beyond the NLSE regime. We can also conclude that the phase convergence combined with the free-wave spectral broadening and consequent bound-wave production at high wavenumbers discussed in Section 4.1 is crucial for crest enhancement of modulated wave trains.

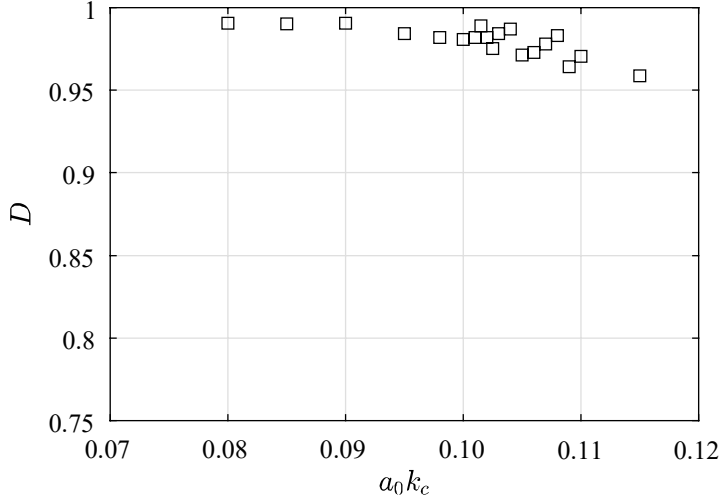

**Figure 13.** Relation between the initial wave steepness and degree of phase convergence at the time of the maximum crest height.

*4.3. Temporal Evolution of Phase Relation among the Carrier and Sideband Waves*

In Section 4.2, we found that $D$ decreases slightly as $a_0 k_c$ increases (Figure 13). This imperfect phase convergence can be attributed to the deviation of the carrier phase from 0 at $t = t_{max}$, which is observed in Figure 11(b). In this section, we interpret this imperfect phase convergence in terms of the temporal variation in the phase relation among the carrier, lower-sideband, and upper-sideband waves in the vicinity of $t = t_{max}$. We also discuss the time sequence of the maximum $D$, maximum $\zeta_{cr}$ (corresponding to $t = t_{max}$), and maximum $A_s$.

During the nonlinear evolution of modulated wave trains, the carrier wave is phase-locked with the sideband waves. In the initial stage of nonlinear evolution, the phases of the carrier and two sideband components need to satisfy the following relation such that the sideband waves grow exponentially [41-43]:

$$\Psi \equiv 2\varphi_c - \varphi_+ - \varphi_- = const., \qquad (12)$$

where $\varphi, \varphi_+,$ and $\varphi_-$ denote the phases of the carrier, upper-sideband, and lower-sideband waves, respectively. This phase-locked state with $\Psi = const.$ persists on the time scale of $O((a_0 k_c)^{-2} T_c)$ until the sideband amplitudes become 20–30% of $a_0$, although $\Psi = const.$ is analytically derived assuming that the sideband amplitudes are infinitesimally small compared with the carrier amplitude [41-43]. We here focus on the behavior of $\Psi$ in the vicinity of the modulation peak.

We first investigate the long-term behavior of $\Psi$ for the modulated wave train. Figure 14 presents the temporal evolution of $\Psi$ together with the amplitude evolutions of the carrier, lower sideband, and upper sideband of modulated wave trains with $a_0 k_c = 0.08$ and $0.105$ as examples. Almost a full recurrence cycle is observed for both $a_0 k_c = 0.08$ and $0.105$, although the recurrence period is much shorter for $a_0 k_c = 0.105$ than for $a_0 k_c = 0.08$. Contrary to the theoretical prediction [Eq. (12)], $\Psi$ varies in time at the initial stage for both cases. This initial behavior is due to the nonlinear wave initialization of the HOSM simulation explained in Section 2.1. For $a_0 k_c = 0.08$, $\Psi$ remains constant at approximately $\pi/2$ after the end of the nonlinear wave initialization and then changes rapidly to $-\pi/2$ near the modulation peak (at approximately $t/T_c = 162$). For $a_0k_c = 0.105$, $\Psi$ varies slowly after the end of the nonlinear wave initialization and, as is the case with $a_0k_c = 0.08$, changes rapidly near the modulation peak (at approximately $t/T_c = 104$). However, its rate of change near the modulation peak is much faster than for $a_0k_c = 0.08$. The time when $\Psi = 0$ almost corresponds to that when the carrier amplitude reaches its minimum and when the lower-sideband amplitude reaches its maximum for both cases, and when the upper-sideband amplitude reaches its maximum only for $a_0 = 0.08$. The constant $\Psi$ at $\pi/2$ just after the nonlinear-wave initialization with $a_0k_c = 0.08$ indicates that $\delta k/k_c = 1/7$ is the most unstable modulated wavenumber. This gives the highest initial growth rate of the sidebands for an initial wave steepness of 0.08 [42]. In addition, the slow variation in $\Psi$ just after the nonlinear-wave initialization for $a_0k_c = 0.105$ implies that the sidebands have imperfect exponential growth due to the nonlinear interaction with wave components other than the initial three waves.

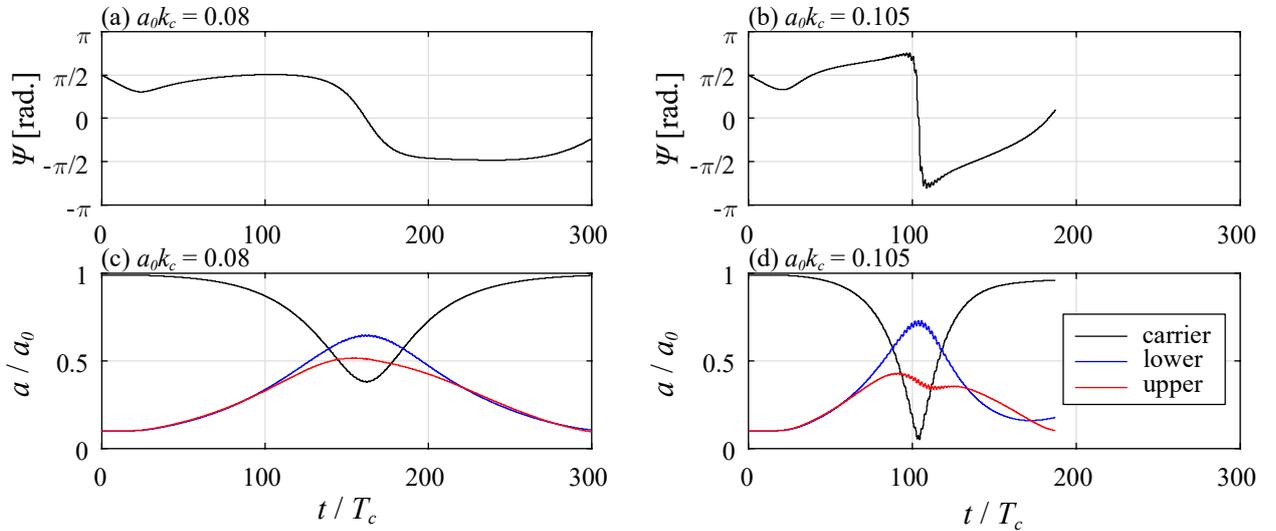

**Figure 14.** Long-term temporal evolutions of (a, b) $\Psi$ and (c, d) the amplitudes of the carrier, lower-sideband, and upper-sideband components of the modulated wave train with $a_0k_c = 0.08$ (a, c) and 0.105 (b, d).

Considering $\varphi_-$ and $\varphi_+$ are almost 0 at $t = t_{max}$ (Figure 11), $\varphi_c$ needs to be 0 for perfect phase convergence. Consequently, $\Psi$ also needs to be 0 at $t = t_{max}$. To investigate $\Psi$ at $t = t_{max}$, we present the temporal evolutions of $\Psi$ for the modulated wave trains with $a_0k_c = 0.08$ and 0.105 at approximately $t = t_{max}$ in Figure 15. At $t = t_{max}$, $\Psi$ is close to 0 ($\Psi = -0.078\pi$) for $a_0k_c = 0.08$ and far from 0 ($\Psi = -0.68\pi$) for $a_0k_c = 0.105$. The value of $\Psi$ at $t = t_{max}$ is related to the difference in time between $\Psi = 0$ and $t = t_{max}$. In both cases, the time at which $\Psi = 0$ is not coincident with $t = t_{max}$. $\Psi$ becomes 0 approximately three wave periods ($3T_c$) ahead of $t = t_{max}$. In addition, the variational speed of $\Psi$ ($d\Psi/dt$) is significantly different between the two cases, as already indicated in Figure 14(a, b). The variational speed with $a_0k_c = 0.08$ is much slower than for $a_0k_c = 0.105$. Therefore, owing to the fast variation in $\Psi$ and the time lag between $\Psi = 0$ and $t = t_{max}$, $\Psi$ at $t = t_{max}$ becomes far from 0 with $a_0k_c = 0.105$.

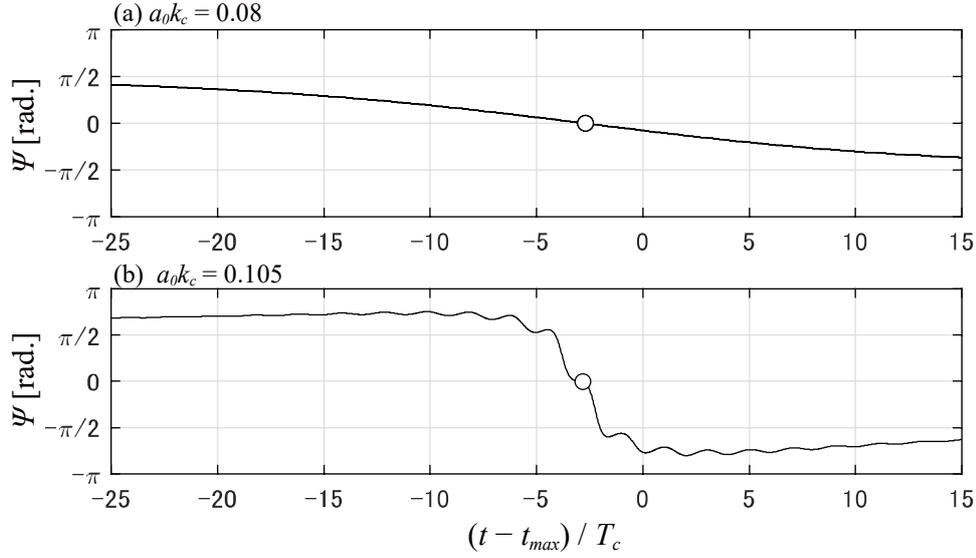

**Figure 15.** Temporal evolution of $\Psi$ for the modulated wave train at approximately the time of the maximum crest height ($t = t_{max}$) for (a) $a_0 k_c = 0.08$ and (b) $a_0 k_c = 0.105$. Circles indicate the times at which $\Psi = 0$.

From the observation above, we conjecture that $\Psi$ becomes farther from 0 at $t = t_{max}$ as $a_0 k_c$ increases because of the faster temporal variation in $\Psi$. To demonstrate this conjecture, we plot $\Psi$ together with its component phases ($\varphi_c, \varphi_-,$ and $\varphi_+$) against $a_0 k_c$ at $t = t_{max}$ in Figure 16. As conjectured, $\Psi$ at $t = t_{max}$ becomes farther from 0 as $a_0 k_c$ increases. Meanwhile, $\varphi_-$ and $\varphi_+$ are confirmed to be close to 0 regardless of $a_0 k_c$. Accordingly, $\varphi_c$ becomes farther from zero at $t = t_{max}$ as $a_0 k_c$ increases. Thus, the slight decrease in degree of phase convergence $D$ at $t = t_{max}$ with increasing $a_0 k_c$ (Figure 13) can be attributed to $\varphi_c$ being out of phase with $\varphi_-$ and $\varphi_+$. The evolution of the phase relation among the carrier, lower-sideband, and upper-sideband waves is found to affect the degree of phase convergence at $t = t_{max}$.

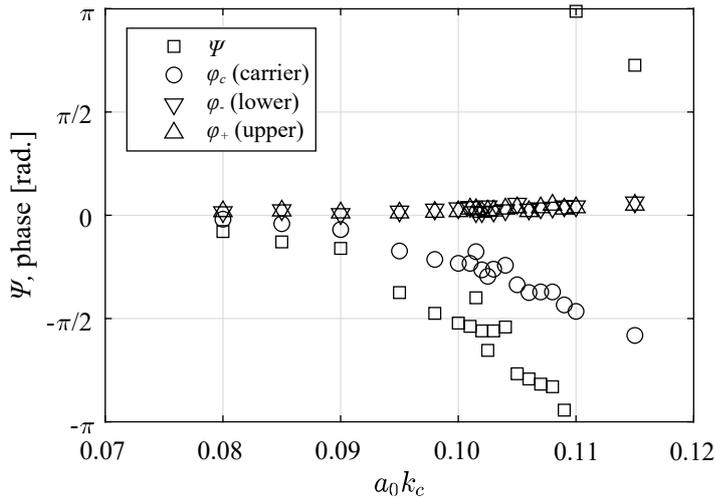

**Figure 16.** Variations of $\Psi$ and the carrier, lower-sideband, and upper-sideband phases at the time of maximum crest height against initial wave steepness $a_0 k_c$.

Lastly, we discuss the sequence of times at which $t = t_{max}$, $\Psi = 0$, and $D$ and $A_s$ reach their maxima. These are plotted against $a_0 k_c$ in Figure 17. The following features, which have been found so far for modulated wave trains with specific $a_0 k_c$ values, seem to be robust regardless of $a_0 k_c$: $A_s$ reaches its maximum about $3T_c$ to $5T_c$ after $t = t_{max}$, $D$ reaches its maximum about $2T_c$ ahead of $t = t_{max}$, and $\Psi$ is 0 approximately $2T_c$ to $4T_c$ ahead of $t = t_{max}$. The close times of maximum $D$ and $\Psi = 0$ indicate that $D$ reaches its maximum when the carrier, lower-sideband, and upper-sideband waves are in phase because the contribution of these three components to $D$ is dominant. Figure 17 also reveals that the time of maximum $A_s$ necessarily lags behind that of maximum $D$ regardless of $a_0 k_c$. The degree of phase convergence $D$ and the potential maximum crest height $A_s$ are the two most significant factors determining the actual maximum crest height. Consequently, we can surmise that the crest height reaches its maximum ($t = t_{max}$) midway between the times of maximum $D$ and maximum $A_s$.

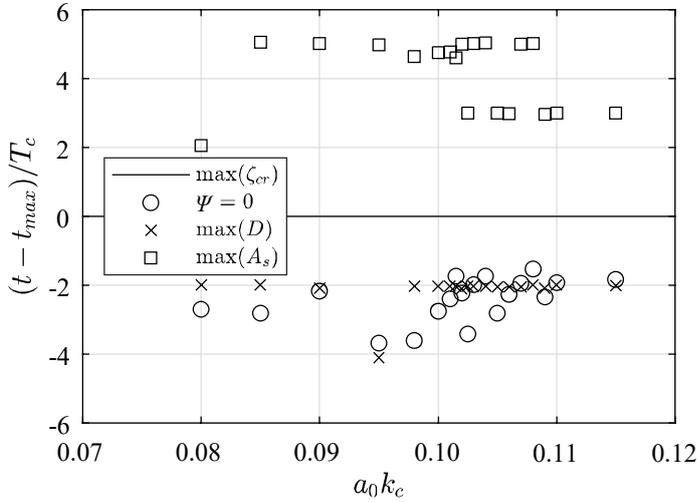

**Figure 17.** Variations in times of maximum crest height ($\zeta_{cr}$), $\Psi = 0$, maximum degree of phase convergence ($D$), and maximum amplitude sum ($A_s$) against initial wave steepness $a_0 k_c$.

## 5. Conclusions

The most notable finding of this study is that the phases of all the spectral wave components of the nonlinearly evolving modulated wave trains coincided at the peak of the modulation. This phase convergence process contributes to the crest enhancement of modulated wave trains beyond the AB solution of the cubic NLSE. However, this phase convergence is a nonlinear process where the phases change in time and is not a linear dispersive focusing where the initial phases are stationary. This was unraveled by numerical examinations based on HOSM up to the fifth order in a non-breaking potential-flow regime. HOSM allowed us to investigate nonlinearity and spectral bandwidth beyond the NLSE regime. The scrutinization of the HOSM outputs also revealed two other critical physical processes of such crest enhancement: spectral broadening and bound-wave production. The free wave spectra of modulated wave trains in the HOSM simulation broaden beyond the AB solution because of the unrestricted spectral bandwidth in the HOSM simulation. The free-wave spectral broadening energizes the bound-wave production at high wavenumbers. The bound wave components can contribute more than a quarter of the maximum crest height at an initial wave steepness of 0.115.

The finding regarding the phase convergence implies that we may not be able to distinguish whether an observed rogue wave was generated due to linear focusing or modulational instability only from a single point measurement. A recent study by Gemmrich and Cicon [44] explained the observed rogue wave by the superposition of linear waves with fourth-order Stokes wave correction. However, based on our study, it seems worthwhile paying more attention to the evolutionary process leading to rogue wave formation.

Finally, we have elucidated the significance of the strong nonlinearity in the evolution of the modulated wave train. When wave-breaking occurs, the maximum crest in the tank falls below the HOSM simulation. However, with a sufficiently large initial wave steepness but without wave breaking, the highest crest in the tank exceeds the HOSM simulation due to a strong nonlinearity. The first finding is not surprising, but the second finding highlights the significance of the strong nonlinearity and warrants further study.

**Author Contributions:** Conceptualization, H.H. and T.W.; methodology, H.H. and H.S.; software, H.H. and T.W.; validation, H.H.; formal analysis, H.H.; investigation, H.H., H.S., and T.W.; resources, H.H. and H.S.; data curation, H.H.; writing—original draft preparation, H.H.; writing—review and editing, H.H. and T.W.; visualization, H.H.; supervision, T.W.; project administration, H.H.; funding acquisition, H.H. and T.W. All authors have read and agreed to the published version of the manuscript.

**Funding:** This research was funded by JSPS KAKENHI grant numbers JP 16H02429, JP21H01538, and JP22H01136.

**Acknowledgments:** We thank Katsuji Tanizawa, Alessandro Iafrati, and Harry Bingham for useful discussions. Mark Kurban from Edanz (https://www.jp.edanz.com/ac) edited a draft of this paper.

**Conflicts of Interest:** The authors declare no conflict of interest.

**Appendix A. Akhmediev Breather Solution**

The Akhmediev breather solution of the NLSE in deep water reads [16]

$$B(x,t) = a_0 \exp\left\{-\frac{i}{2}\epsilon^2 \omega_c t\right\} \left[\frac{\rho \cosh(\Omega t) - i\gamma \sinh(\Omega t)}{\cosh(\Omega t) - \sqrt{1-\rho^2/2}\cos\{k_c(x-c_{gc}t)/N_k\}} - 1\right] \quad (13)$$

with

$$\epsilon = a_0 k_c, c_{gc} = \frac{1}{2}\frac{\omega_c}{k_c}, \rho = \frac{1}{2\epsilon N_k}, \gamma = \frac{\sqrt{2}}{2\epsilon N_k}\sqrt{1-\left(1-\frac{1}{2\sqrt{2}\epsilon N_k}\right)^2}, \quad (14)$$

$$\Omega = \frac{1}{2}\gamma\epsilon^2\omega_c, N_k = \frac{k_c}{\delta k},$$

where $B(x,t)$ is the complex amplitude of the surface elevation $\zeta(x,t)$. The term $\hat{\delta}$ defined in Eq. (4) corresponds to the inverse of $\epsilon N_k$:

$$\hat{\delta} = \frac{1}{2\epsilon N_k}. \quad (15)$$

Within the framework of the deep-water NLSE, $\zeta(x,t)$ can be expressed as follows considering the nonlinearity up to the second order [45]:

$$\zeta(x,t)\left(\equiv \zeta^{(f+b)}(x,t)\right) = \zeta^{(f)}(x,t) + \zeta^{(b)}(x,t) \quad (16)$$

with

$$\begin{cases} \zeta^{(f)}(x,t) = Re[B \exp\{i(k_c x - \omega_c t)\}], \\ \zeta^{(b)}(x,t) = Re[k_c B^2 \exp\{2i(k_c x - \omega_t)\}], \end{cases} \quad (17)$$

where $\zeta^{(f)}(x,t)$ and $\zeta^{(b)}(x,t)$ represent the free wave and second-order bound wave, respectively.

The maximum amplitude, which corresponds to the maximum crest height, of the free wave is attained at $(x,t) = (0,0)$ and reads [14-16,21]

$$\frac{\max(\zeta^{(f)})}{a_0} = 1 + 2\sqrt{1 - \frac{1}{2}\hat{\delta}^2}. \quad (18)$$

From Eqs. (16)–(18), the maximum crest height considering the contribution of the second-order bound wave can be expressed as

$$\frac{\max(\zeta^{(f+b)})}{a_0} = \frac{\max(\zeta^{(f)})}{a_0}\left\{1 + \frac{1}{2}k_c \max(\zeta^{(f)})\right\}. \quad (19)$$

**Appendix B. Phase of Second-Order Superharmonic and Subharmonic Waves**

In this section, we investigate the relation between the phases of the free and second-order bound waves. We address the second-order bound waves produced by a pair of free waves propagating in the positive $x$ direction. The free-surface elevation $\zeta^{(1)}$ consisting of a pair of free waves with wavenumbers $k_1$ and $k_2$ is expressed as

$$\zeta^{(1)}(x,t) = \sum_{j=1}^{2} |\hat{\zeta}_j| \cos \theta_j \tag{20}$$

with

$$\theta_j = k_j x - \omega_j t + \gamma_j, \tag{21}$$

where $j (= 1, 2)$ and $\gamma$ denote the indexes of the free waves and the initial phase, respectively. We assume deep water and $k_1 \geq k_2 \geq 0$. The second-order bound waves $\zeta^{(2)}$ produced by the pair of free waves [Eqs. (20) and (21)] are expressed as [46]

$$\zeta^{(2)}(x,t) = \sum_{j=1}^{2} \frac{|\hat{\zeta}_j|^2 k_j}{2} \cos(2\theta_j) + |\hat{\zeta}_1||\hat{\zeta}_2| \frac{k_1 + k_2}{2} \cos(\theta_1 + \theta_2) \\ + |\hat{\zeta}_1||\hat{\zeta}_2| \frac{-(k_1 - k_2)}{2} \cos(\theta_1 - \theta_2). \tag{22}$$

The first two terms represent superharmonics, and the last term represents subharmonics.

We assume that the free waves are in phase at a specific time ($t = t_0$) and location ($x = x_0$); that is, $\theta_1 = \theta_2 = 0$. Then, the free and the second-order bound-wave solution reads

$$\begin{cases} \zeta^{(1)}(x_0, t_0) = \sum_{j=1}^{2} |\hat{\zeta}_j|, \\ \zeta^{(2)}(x_0, t_0) = \sum_{j=1}^{2} \frac{|\hat{\zeta}_j|^2 k_j}{2} + |\hat{\zeta}_1||\hat{\zeta}_2| \frac{k_1 + k_2}{2} + |\hat{\zeta}_1||\hat{\zeta}_2| \frac{-(k_1 - k_2)}{2}. \end{cases} \tag{23}$$

Only the second-order subharmonic component is negative, although the free-wave and second-order superharmonic components are positive. Therefore, when all the free-wave components are in phase, the second-order superharmonics are in phase, and subharmonics are in counter-phase with the free waves.